\documentclass[twocolumn, showkeys, newabstract]{revtex4}
\usepackage{epsfig}

\begin{document}

\title{Kinetic model of DNA replication in eukaryotic organisms}

\author{John Herrick${}^1$, Suckjoon Jun${}^2$, John Bechhoefer${}^{2*}$, 
Aaron Bensimon${}^{1*}$}

\affiliation{${}^1$Unit\'e de Stabilit\'e des G\'enomes, D\'epartement Structure et Dynamique des G\'enomes , Institut Pasteur, 25-28, rue du Dr. Roux, 75724 
Paris Cedex 15, France}

\affiliation{${}^2$Department of Physics, Simon Fraser University, 
Burnaby, British Columbia, V5A 1S6, Canada}

\affiliation{${}^*$Corresponding authors.  E-mail:
johnb@sfu.ca (J.B.) and abensim@pasteur.fr (A.B.)}

\begin{abstract}
We formulate a kinetic model of DNA replication
that quantitatively describes recent results on DNA replication 
in the {\it in vitro} system of {\it Xenopus laevis} prior to the 
mid-blastula transition.  The model describes well a large
amount of different data within a simple theoretical framework.
This allows one, for the first time, to determine the parameters 
governing the DNA replication program in a eukaryote on a 
genome-wide basis. In particular, we have determined the frequency 
of origin activation in time and space during the cell cycle. 
Although we focus on a specific stage of development, this model 
can easily be adapted to describe replication in many other organisms, 
including budding yeast.\\

\noindent Journal Ref: {\it J.Mol.Biol.}, {\bf 320}, 741-750 (2002) 
\end{abstract}


\maketitle

\section*{Introduction}

        Although the organization of the genome for DNA replication 
varies considerably from species to species, the duplication of most 
eukaryotic genomes shares a number of common features: \vspace*{1em}

\noindent 
1) DNA is organized into a sequential series of replication 
units, or replicons, each of which contains a single origin of 
replication.\cite{Hand, Friedman}  \vspace*{1em}

\noindent 
2)  Each origin is activated not more than once during the 
cell-division cycle.  \vspace*{1em}

\noindent 
3) DNA synthesis propagates at replication forks 
bidirectionally from each origin.\cite{Cairns}   \vspace*{1em}

\noindent 
4) DNA synthesis stops when two newly replicated regions of DNA meet.  
\vspace*{1em}

Understanding how these parameters are coordinated during the 
replication of the genome is essential for elucidating the mechanism
by which $S$-phase is regulated in eukaryotic cells.
In this article, we formulate a stochastic model based on these 
observations that yields a mathematical
description of the process of DNA replication and provides a convenient 
way to use the full statistics gathered in any particular replication 
experiment.  It allows one to deduce accurate values for the parameters 
that regulate DNA replication in the {\it Xenopus laevis}
replication system, and it can be generalized 
to describe replication in any other eukaryotic system. This type of model
has also been shown to apply for the case of RecA polymerizing on a single
molecule  of DNA.\cite{Shivashankar}  The model, as described 
in the methods section below, turns out to be 
formally equivalent to a well-known stochastic description of the 
kinetics of crystal growth, which allows us to draw on a number of 
previously derived results and, perhaps equally 
important, suggests a vocabulary that we find useful and 
intuitive for understanding the process of replication.

%\section*{Crystal-Growth Kinetics and DNA Replication}

Since the kinetics of DNA replication in any cell system depends on two 
fundamental quantities, replication fork velocity and initiation 
frequency, one of the principal goals of this kind of analysis is to 
derive accurate values for these quantities, including
any variation, during the course of $S$-phase.  As
replicon size and the duration of $S$-phase depend on the
values of these parameters, this information
is indispensable for understanding the mechanisms regulating $S$-phase 
in any given cell system.\cite{Pierron, Walter_Newport, Hyrien_Mechali, Coverly_Laskey, Blow_Chong, Shinomiya, Brewer_Fangman, Gomez}

\section*{Results}

\subsection*{Summary of the {\it X. Laevis} replication experiment}

Here, we describe recent experimental results obtained on the kinetics 
of DNA replication in the well-characterized {\it Xenopus laevis} 
cell-free system.\cite{Herrick_JMB, Lucas}  One of the main goals of this 
paper will be to show that using the theoretical approach described below, 
one can extract more information -- and more reliably -- than before 
from such experiments.

\begin{figure}[ht]
\centering
\epsfig{file=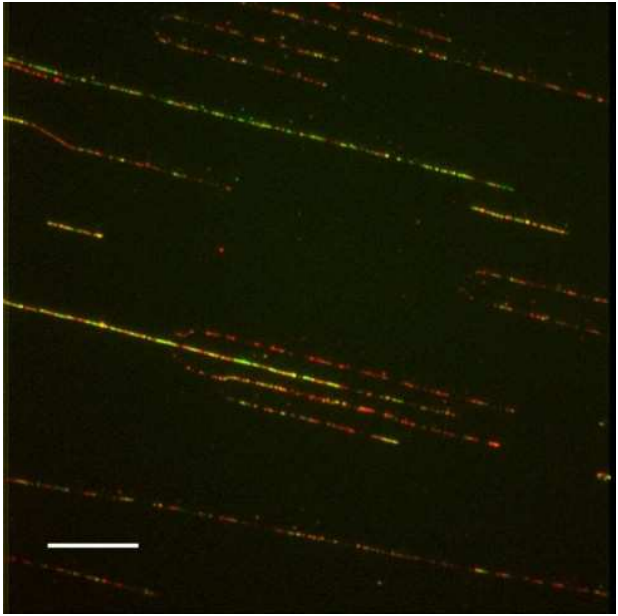, width=3.2in}
\caption{A fluorescence micrograph (bar $=$ 20 $\mu$m).  Early replicating sequences labeled with biotin-dUTP are visualized using red fluorescing antibodies (Texas Red).  Later replicating sequences are in addition labeled with dig-dUTP and visualized using green (FITC) fluorescing antibodies.}
\label{fig:fluorescence}
\end{figure}

\begin{figure}[ht]
\centering
\epsfig{file=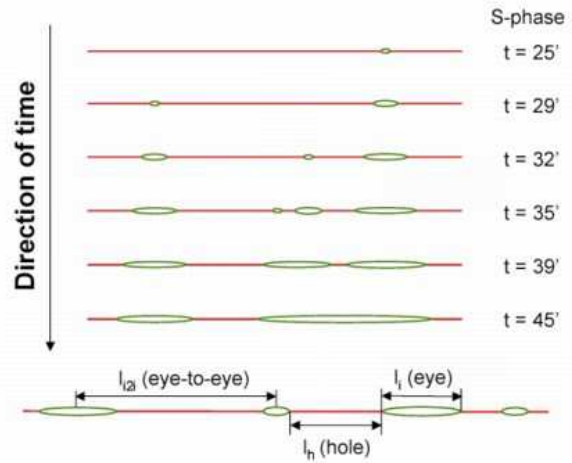, width=3.2in}
\caption{Schematic representation of labeled and combed DNA 
molecules. Since replication initiates at multiple dispersed sites 
throughout the genome, the DNA can be differentially labeled, so 
that each linearized molecule contains alternating 
subregions stained with either one or both dyes.  The bubbles correspond to sequences synthesized in the presence of 
a single dye (red). The green segments correspond to those 
sequences that were synthesized after the second dye (green) was added. The result is 
an unambiguous distinction between eyes and holes (earlier and 
later replicating sequences) along the linearized molecules. Replication 
is assumed to have begun at the midpoints of the bubble sequences and to have proceded bidirectionally from the site where DNA synthesis was initiated. 
Measurements between the centers of adjacent eyes 
provide information about replicon sizes (eye-to-eye distances). The fraction 
of the molecule already replicated by a given time, $f(\tau)$, is determined 
by summing the lengths of the bubbles and 
dividing that by the total length of the respective molecule.}
\label{fig:model}
\end{figure}

In the {\it Xenopus} replication experiments, fragments of DNA 
that have completed one cycle of 
replication are stretched out on a glass surface using molecular combing.\cite{Bensimon, Michalet, Herrick_PNAS}  Typical two-color epifluorescence images of the combed DNA are shown in Fig. \ref{fig:fluorescence}. The DNA that has replicated 
prior to some chosen time $t$ is labeled with a single fluorescent dye, 
while DNA that replicated after that time is labeled with two dyes. 
The result is a series of samples, each of which 
corresponds to a different time $t$ during $S$-phase. 
Using an optical microscope, one can directly
measure eye, hole, and eye-to-eye lengths at that time. 
We can thus monitor the evolution of genome duplication from time point to 
time point, as DNA synthesis advances.  
(See Fig. \ref{fig:model}.)

Cell-free extracts of eggs from {\it Xenopus laevis} support the major
transitions of the eukaryotic cell cycle, including complete chromosome
replication under normal cell-cycle control and offers the opportunity
to study the way that DNA replication is coordinated within the cell
cycle.  In the experiment, 
cell extract was added at $t =$ 2', and $S$-phase began 15 to 
20' later.  DNA replication was monitored by incorporating two different 
fluorescent dyes into the newly synthesized DNA.  The first dye was 
added before the cell enters $S$-phase in order to label the entire 
genome.  The second dye was added at successive time points $t =$ 25,
29, 32, 35, 39, and 45', in order to label the later replicating 
DNA.  DNA taken from each time point was combed, and measurements were made 
on replicated and unreplicated regions.  
The experimental details are described elsewhere\cite{Herrick_JMB}, 
but the approach is similar to
DNA fiber autoradiography, a method that has been in use for the 
last 30 years.\cite{Huberman, Jasny}  Indeed, the same approach has recently been adapted to study the regulatory 
parameters of DNA replication in HeLa cells.\cite{Jackson_Pombo}  Molecular combing, however, has the advantage that a	
large amount of DNA may be extended and aligned on a glass slide
which ensures significantly better statistics (over several thousand 
measurements corresponding to several hundred genomes per coverslip).  
Indeed, the molecular combing experiments provide, for the first time, 
easy access to the quantities of data necessary for testing models 
such as the one advanced in this paper.

\subsection*{Generalization of the model to account for specific 
features of the {\it X. laevis} experiment}

The experimental results obtained on the kinetics of DNA 
replication in the {\it in vitro} cell-free system of {\it Xenopus 
laevis} \cite{Herrick_JMB, Lucas} were analyzed 
using the kinetic model developed below.  In formulating that model, 
we found that we had to take into account explicitly a number of 
observations that are peculiar to the particular experiment analyzed:

\vspace*{1em}

\noindent 
1)  One goal of the experiment is to measure the initiation function 
$I(\tau)$, which is the probability of initiating an origin at time $\tau$,
per unit length of unreplicated DNA.  The simplest assumptions, in 
terms of our model, would be that either $I$ is peaked at or near 
$\tau=0$ (all origins initiated at the beginning of $S$-phase) or 
$I(\tau) = $ constant, (origins initiated at constant rate throughout 
$S$-phase).  However, neither assumption turns out to be consistent 
with the data analyzed here; thus, we formulated our model to allow 
for arbitrary initiation patterns and deduced an estimate for 
$I(\tau)$ directly from the 
data.  We note that initiation is believed 
to occur synchronously during the first half of $S$-phase in
{\it Drosophila melanogaster} early embryos.\cite{Shinomiya, Blumenthal}  Initiation in the myxomycete {\it Physarum polycephalum,} on the other 
hand, occurs in a very broad temporal window, suggesting that 
initiation occurs 
continuously throughout $S$-phase.\cite{Pierron} 
Finally, recent observations suggest that in {\it Xenopus laevis}, early 
embryos nucleation may occur with increasing frequency as DNA synthesis 
advances.\cite{Herrick_JMB, Lucas}  
By choosing an appropriate
form for $I(\tau)$, one can account for any of these scenarios.  
Below, we show how measured quantities may, using the model, 
be inverted to provide an estimate for $I(\tau)$. \vspace*{1em}

\noindent 
2)  The basic form of the model assumes implicitly that the DNA 
analyzed began replication at $\tau = 0$, but this may not be so, for two 
reasons:   \vspace*{1em}

i)  In the experimental protocols, the DNA 
analyzed comes from approximately 20,000 independently 
replicating nuclei.  Before each genome can replicate, its nuclear 
membrane must form, along with, presumably, the replication factories. 
This process takes 15-20 minutes.\cite{Blow_Laskey, Blow_Watson, Wu}  Because the exact amount of time can vary from cell to cell, 
the DNA analyzed at time $t$ in the laboratory may have started 
replicating over a relatively wide range of times.  \vspace*{1em}

ii)  In eukaryotic organisms, origin activation may be distributed
in a programmed manner throughout the length of $S$-phase, and, as a
consequence, each origin is turned on at a specific time (early and 
late).\cite{Simon}  \vspace*{1em}

\noindent 
In the current experiment, the lack of information about the 
locations of the measured DNA segments along the genome means that we 
cannot distinguish between asynchrony due to reasons (i) or 
(ii).  We can however account for their combined effects by
introducing a starting-time distribution $\phi(t')$, which is 
the probability---for whatever reason---that a given piece of analyzed 
DNA began replicating at time $t'$ in the lab.  Using our model, we 
can directly extract the starting time distribution from the data.
\vspace*{1em}

\noindent 
3)  The models described above assumed that statistics could be 
calculated on infinitely long segments of DNA.  In the experimental 
approach, the combed DNA is broken down into relatively short segments 
(100 kb, typically).  Although it is difficult to account for this effect 
analytically, we wrote a Monte-Carlo simulation that can mimic 
such ``finite-size'' effects.\vspace*{1em}

\noindent 
4)  The experiments are all analyzed using an epifluorescence 
microscope to visualize the fluorescent tracks of combed DNA on glass slides.  
The spatial resolution ($\approx$ 0.3 $\mu$m) means that smaller signals 
will not be detectable.  Thus, two replicated segments separated by an 
unreplicated region of size $<$ 0.3 $\mu$m will be falsely assumed 
to be one longer replicated segment.  We accounted for this in 
the Monte-Carlo simulations by calculating statistics on a coarse lattice whose 
size equalled the optical resolution, while the simulation itself takes 
place on a finer lattice.

\subsection*{Application of the kinetic model to the analysis of 
DNA replication in {\it X. Laevis}}

\begin{figure}[ht]
\centering
\epsfig{file=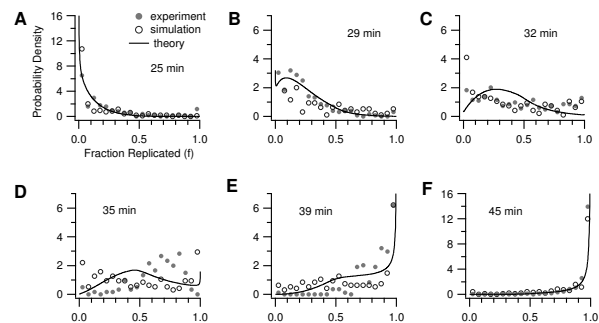, width=3.2in}
\caption{$\rho(f,t)$ distributions for the 6 time points. The curves show the probability that a molecule at a given time point (A-F) has undergone a certain amount of replication before the second dye was added. The filled circles represent the experimental data. The results of the Monte-Carlo simulation are shown in open circles; analytical curves are the global fitting.}
\label{fig:rho}
\end{figure}

Using the generalizations discussed above, we analyzed recent results 
obtained on DNA replication in the {\it Xenopus laevis} cell-free system.
DNA taken from each time point was combed, and measurements were made 
on replicated and unreplicated regions.  Statistics from each time 
point were then compiled into six histograms (one for each time point)
of the distribution $\rho(f,t)$ of replicated fractions $f$ at time $t$ 
(Fig. \ref{fig:rho}).

\begin{figure}[ht]
\centering
\epsfig{file=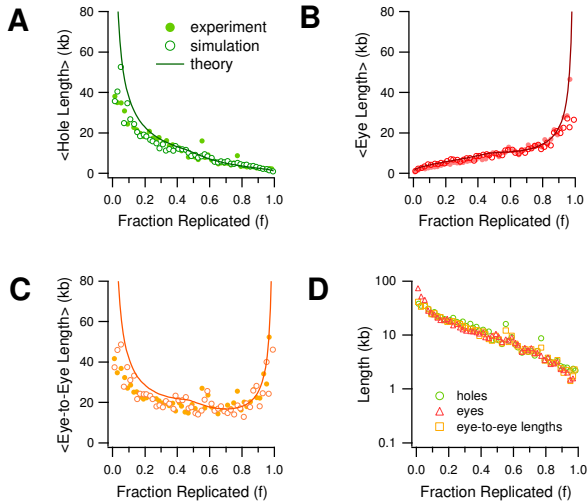, width=3.2in}
\caption{Mean quantities vs. replication fraction.  {\bf (A)} average hole size $\ell_h(f)$; {\bf (B)} average eye size $\ell_i(f)$; {\bf (C)} average eye-to-eye size $\ell_{i2i}(f)$.  Filled circles are data; open circles are from the Monte-Carlo simulation; the solid curve is a least-squares fit, based on a two-segment $I(\tau)$; {\bf (D)} curves in {\bf (A)-(C)} collapsed onto a single plot, confirming mean-field hypothesis.  (The discrepancies near $f = 0$ and $1$ reflect measurement errors.  Very small eyes or holes may be missed because of limited optical resolution; very large eyes or holes may be eliminated because of finite segment sizes.)} 
\label{fig:mean-curves}
\end{figure}

One can immediately see from Fig.~\ref{fig:rho} the need 
to account for the spread in starting times.  If all the segments of 
DNA that were analyzed had started replicating at the same time, then 
the distributions would have been concentrated over a very small 
range of $f$.  But, as one can see in Fig. \ref{fig:rho}C, some 
segments of DNA (within the same time point) have already finished 
replicating ($f = 1$) before others have even started ($f = 0$).  
This spread is far larger than would be expected on account of the 
finite length of the segments analyzed.
Because of the need to account for the spread in starting times, it is
simpler to begin by sorting data by the replicated fraction $f$ of the
measured segment.  We thus assume that all segments with a similar 
fraction $f$ are at roughly the same point in $S$-phase, an 
assumption that we can check by partitioning the data into subsets and 
redoing our measurements on the subsets.  In Fig. \ref{fig:mean-curves}A-C, we
plot the mean values $\ell_h$, $\ell_i$, and $\ell_{i2i}$ against $f$.

\begin{figure}[ht]
\centering
\epsfig{file=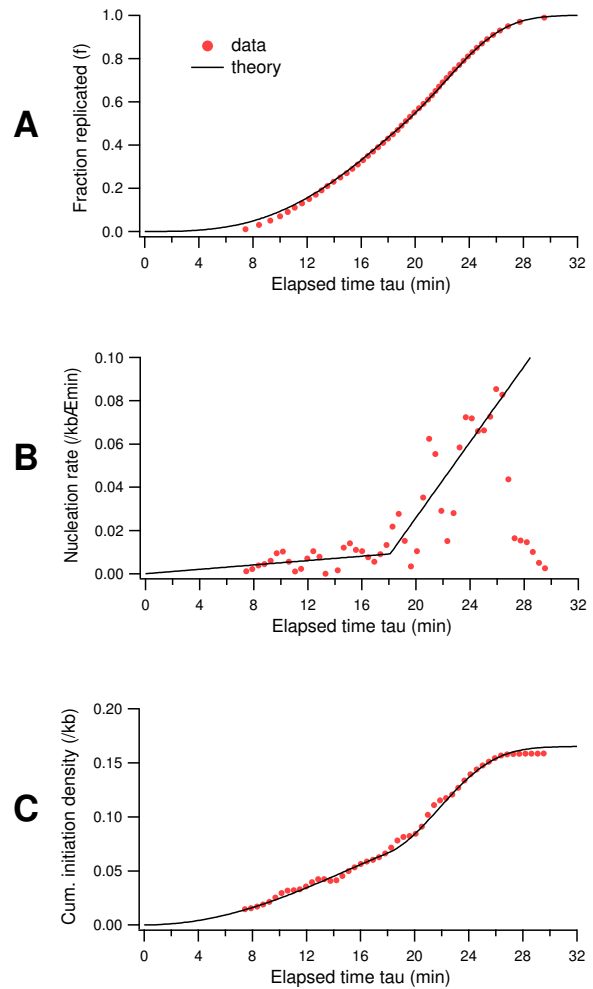, width=3.2in}
\caption{{\bf (A)}  Fraction of replication completed, $f(\tau)$.  Red points are derived from the measurements of mean hole, eye, and eye-to-eye lengths.  Black curve is an analytic fit (see below).  {\bf (B)}  Initiation rate $I(\tau)$.  The large statistical scatter arises because the data points are obtained by taking two numerical derivatives of the $f(\tau)$ points in A.  {\bf (C)}  Integrated origin separation, $I_{tot}(\tau)$, which gives the average distance between all origins activated up to time $\tau$.  In A-C, the black curves are from fits that assume that $I(\tau)$ has two linear regimes of different slopes.  The form we chose for $I(\tau)$ was the simplest analytic form consistent with the data in B.  The parameters for the least-squares fits (slopes $I_1$ and $I_2$, break point $\tau_1$) are obtained from a global fit to the eight data sets in Fig. \ref{fig:rho}A-F and Fig. \ref{fig:mean-curves}A-B, {\it i.e.}, $\rho(f)$ from six time points, $\ell_h(f)$, and $\ell_i(f)$.}
\label{fig:misc}
\end{figure}

We then find $f(\tau)$, $I(\tau)$, and the
cumulative distribution of lengths between activated origins of 
replication, $I_{tot}(\tau)$.  (See Fig. \ref{fig:misc}.)
The direct inversion for $I(\tau)$ (Fig. \ref{fig:misc}B)
shows several surprising features:  First, origin activation 
takes place throughout $S$-phase and 
with increasing probability (measured relative to the amount of
unreplicated DNA), as recently inferred from a cruder analysis
of data from the same system using plasmid DNA.\cite{Lucas}  
Second, about halfway through $S$-phase, there is
a marked increase in initiation rate, an observation that, if 
confirmed, would have biological significance.
It is not known what might cause a sudden increase (break point) 
in initiation frequency halfway through $S$-phase. The increase 
could reflect a change in chromatin structure 
that may occur after a given fraction of the genome has undergone 
replication. This in turn may increase the number of 
potential origins as DNA synthesis advances.\cite{Pasero} 

The smooth curves in Fig \ref{fig:mean-curves}A-C are fits based on the model, using
an $I(\tau)$ that has two linearly increasing regions, with arbitrary 
slopes and ``break point'' (three free parameters).  The fits are 
quite good, except where the finite size of the combed DNA fragments 
becomes relevant.  For example, when mean hole, eye, and eye-to-eye 
lengths exceed about 10\% of the mean fragment size, larger segments 
in the distribution for $\ell_h(f)$, etc., are excluded and the averages 
are biased down.  We confirmed this with the Monte-Carlo simulations, 
the results of which are overlaid on the experimental data.  The 
finite fragment size in the simulation matches that of the 
experiment, leading to the same downward bias.  In Fig. 
\ref{fig:misc}, we overlay the fits on the experimental data.
We emphasize that we obtain $I(\tau)$ directly from the data,
with no fit parameters, apart from an overall scaling of the time axis.  
The analytical form is just a model that 
summarizes the main features of the origin-initiation rate we determine 
via our model, from the experimental data.  
The important result is $I(\tau)$.  
From the maximum of $I_{tot}(\tau)$,
we find a mean spacing between activated origins of 6.3 $\pm$ 0.3 kb, 
which is much smaller than the minimum mean eye-to-eye separation 14.4 
$\pm$ 1.5 kb.  

\begin{figure}[ht]
\centering
\epsfig{file=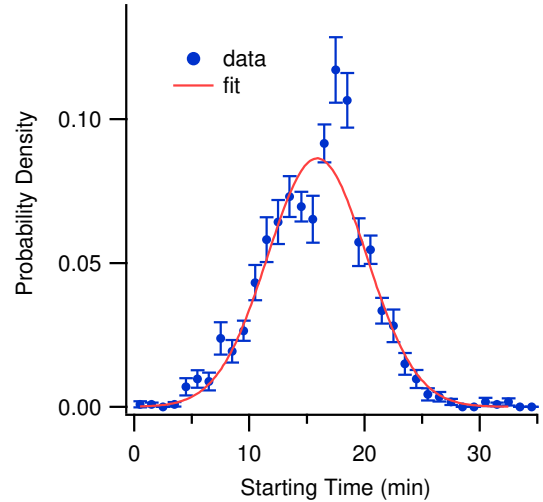, width=3.2in}
\caption{Starting-time distribution $\phi(t)$.  Solid curve is a 
least-squares fit to a Gaussian distribution.}
\label{fig:starting-time}
\end{figure}

In our model, the two quantities differ if initiation takes place throughout S-phase, as coalescence of replicated regions leads to fewer domains, and hence fewer inferred origins (see the note below Eq. \ref{eq:ell-i2i-tau} on p. 16).  The mean eye-to-eye separation is of particular interest because its inverse is just the domain density (number of active domains per length), which can be used to estimate the number of active replication forks  at each moment during $S$-phase.  For example, 
the saturation value of $I_{tot}$ corresponds to the maximum number (about 480,000/genome) of active origins of replication.  Since there are about 400 replication foci/cell nucleus, this 
would indicate a partitioning of approximately 1,200 origins (or, equivalently, about 7.5 Mb) per replication focus.\cite{Blow_Laskey, Mills}
The distribution of $f$ values in the $\rho(f,t)$ plots can be used to deduce the starting-time distribution ($\phi(t')$), along with the fork velocity $v$.  (Fig.~\ref{fig:starting-time}).  The 
spread in starting times $\phi$ is consistent with a Gaussian distribution, with a mean of $15.9 \pm 0.6$ min. and a standard deviation of $6.1 \pm 0.6$ min.  For the fork velocity, we find $v = $ 615 $\pm$ 35 bases/min., in excellent agreement with previous estimates.\cite{Mahbubani, Lu}  As with the $f$ data, we extracted $\phi(t)$ and $v$ from a global fit to data from all 
six time points.

\section*{Discussion}
\subsection*{Initiation throughout $S$-phase}
The view that we are led to here, of random initiation events occurring 
continuously during the replication of {\it Xenopus} sperm chromatin in egg 
extracts, is in striking contrast to what has until recently been the 
accepted view of a regular periodic organization of replication origins 
throughout the genome.\cite{Buongiorno-Nardelli, Laskey, Coverly_Laskey, Blow_Chong}  For a discussion of experiments that raise 
doubts on such a view, see Berezney.\cite{Berezney}
The application of our model to the results of Herrick {\it et al.}
indicates that the kinetics of DNA replication in the {\it X. laevis} 
{\it in vitro} system closely resembles that of genome duplication in early 
embryos.  Specifically, we find that the time required to duplicate 
the genome {\it in vitro} agrees well with what is observed {\it in vivo}. 
In addition, the model yields accurate values 
for replicon sizes and replication fork velocities that confirm 
previous observations.\cite{Mahbubani, Hyrien_Mechali}  Though replication 
{\it in vitro} may differ biologically from what occurs {\it in vivo}, 
the results nevertheless demonstrate that the kinetics remains 
essentially the same.  Of course, the specific finding of an increasing
rate of initiation invites a biological interpretation involving a kind of 
autocatalysis, whereby the replication process itself leads to the 
release of a factor whose concentration determines the rate of 
initiation.  This will be explored in future work.
\subsection*{Directions for future experiments in {\it X. laevis}}
One effect that we did not include in our analysis is a variable
fork velocity.  For example, $v$ might decrease 
as forks coalesce or as replication factor becomes limiting toward the 
end of $S$ phase.\cite{Blow_Laskey, Blow_Watson, Wu, Pierron}  Such effects, if present, are 
too small to see in the data analyzed here.
	
Another important question is to separate the effects of
any intrinsic distribution due to early and late-replicating regions 
of the genome of a single cell from the extrinsic distribution caused
by having many cells in the experiment.  One approach would be to 
isolate and comb the DNA from a {\it single} cell.  
Although difficult, such an experiment is technically feasible.
The latter problem could be resolved by {\it in situ} fluorescence 
observations of the chosen cell.
\subsection*{Applications to other systems}
One can entertain many further applications of the 
basic model discussed above, which can be generalized, if need be.
For example, Blumenthal {\it et al.} interpreted their results on 
replication in {\it Drosophila melanogaster} for 
$\rho_{i2i}(\ell,f)$ to imply periodically spaced origins in the 
genome.\cite{Blumenthal}  (See their Fig. 7.) It is difficult to 
judge whether their peaks are real or statistical happenstance,
but if the conclusion is indeed that the origins in that system are 
arranged periodically, the kinetics model could be generalized in a 
straightforward way (introducing an $I(x,\tau)$ that was periodic in $x$).
	
Very recently, detailed data on the replication of budding 
yeast ({\it Saccharomyces cerevisiae}) have become available.\cite{Raghuraman}  The data provide information on the 
locations of origins and the timings of their initiation during 
$S$-phase.  These data support the view of origin initiation 
throughout $S$-phase.  
Unlike replication in {\it Xenopus} prior to the mid-blastula 
transition, origins in budding yeast are associated with highly 
conserved sequence elements (autonomous replication sequence elements, 
or ARSs).  Raghuraman {\it et al.}\cite{Raghuraman}  also give the first estimates of 
the {\it distribution} of fork velocities during replication.  
Although broad, the distribution is apparently stationary, and there 
is no correlation between velocities and the time in $S$-phase when 
the forks are initiated.  The model developed here could be 
generalized in a straightforward way to the case of budding yeast.  
Knowing the sequence of the genome and hence the location of 
potential origins means that the initiation function
would be an explicit function of position $x$ along the genome, with 
peaks of varying heights at each potential origin.  The advantage of 
the kind of modeling advanced here would be the opportunity to derive 
quantities such as the replication fraction as a function of time in 
$S$-phase.  Raghuraman {\it et al.} fit their data for this ``timing curve''
to an arbitrarily chosen sigmoidal function.  (See their 
supplementary data, Section II-5.)  Such modeling will make it easier 
to find meaningful biological explanations of the programming of 
$S$-phase evolution.

\subsection*{The origin-spacing problem}
	
One outstanding issue in DNA replication in eukaryotes is the observation that the replication origins cannot be too far apart, as this would prevent the genome from being replicated completely within the length of a single $S$-phase.\cite{Gilbert}  One solution that has been proposed is that there is an excess of pre-replication complexes (pre-RCs) of highly conserved proteins, which assemble at ORC-bound DNA sites before the cell enters $S$-phase (e.g., Lucas {\it et al.}\cite{Lucas}, and references therein).  In this case, the position of each potential origin of replication (POR) can be distributed randomly, with a statistically insignificant probability of having large gaps between PORs.  The problem with this solution is that the average POR spacing must be much smaller (less than 1-2 kb) than the reported values of XORC spacing of 7-16 kb.\cite{Walter_Newport, Rowles}
	
A second proposed solution to the origin-spacing problem is to invoke correlations in POR spacings.  In other words, instead of assuming a purely random pre-RC distribution, one imposes constraints that force a partial periodicity on the POR spacing, so that most of the origins are spaced 5-15 kb apart (Blow {\it et al.},\cite{Blow_etal} and references therein).  This suppresses the formation of large gaps but raises other issues.  First, it requires an unknown mechanism to achieve this periodicity of POR spacing.  Second, it assumes implicitly that most of the PORs fire during $S$-phase, to prevent the 30 kb gap that could arise from a originÕs failure to initiate, which is not obvious at all.  Third, if origins initiate throughout $S$-phase, then there needs to be some kind of correlation that forces the more widely spaced origin groups to initiate early enough in $S$-phase to complete replication in the required time.

Implicitly, our model adopts language consistent with the first solution, but it is straightforward to consider the correlations assumed in the second solution.  The presence of significant correlations in PORs would not invalidate the results presented here, which pertain to mean quantities (e.g., Fig. \ref{fig:mean-curves}); however, it would change their interpretation and could change biological models that one might try to make to explain the observed kinetic parameters we extract using the KJMA model.  We plan to investigate these questions, along with the effect of origin efficiency on DNA replication kinetics, in future work.

\section*{Conclusion}

In this article, we have introduced a class of theoretical models
for describing replication kinetics that is inspired by well-known
models of crystal-growth kinetics.  The model allows us to extract the
rate of initiation of new origins, a quantity whose time dependence
has not previously been measured.  With remarkably few parameters, 
the model fits quantitatively the most detailed existing experiment 
on replication in {\it Xenopus}.  It reproduces known results 
(for example, the fork velocity) 
and provides the first reliable description of the temporal 
organization of replication initiation in a higher eukaryote.  
Perhaps most important, the model can be generalized in a 
straightforward way to describe replication and extract relevant
parameters in essentially any organism.

\section*{Methods}

\subsection*{Mathematical analogy between crystal growth and the 
kinetics of DNA replication}
	
In this section, we describe how certain features of the mathematics 
describing crystal growth may be mapped onto a model describing the 
kinetics of DNA replication.  We emphasize that the analogy is a 
formal one -- the underlying processes are completely different.  
However, by mapping our problem onto one that has been long studied 
in a different context, we can take over a number of results that have 
already been derived, and we can develop useful intuitions about 
how to look at experimental results about DNA replication.
	
In the 1930s, several scientists independently derived a 
stochastic model that described the kinetics of crystal growth.\cite{Kolmogorov, Johnson_Mehl, Avrami} The ``Kolmogorov-Johnson-Mehl-Avrami'' (KJMA) model has since been widely 
used by metallurgists and other scientists to analyze thermodynamic phase transformations.\cite{Christian}
	
In the KJMA model, freezing kinetics result from three simultaneous processes: 

\noindent 1) nucleation, which leads to discrete solid domains.  \vspace*{1em}

\noindent 2) growth of the domain.  \vspace*{1em}

\noindent 3) coalescence, which occurs when two expanding domains merge.
\vspace*{1em}
	Each of these processes has an analog in DNA replication in higher eukaryotes, and more specifically embryos:
\noindent 1) The activation of an origin of replication is analogous to the 
nucleation of the solid domains during crystal growth.  \vspace*{1em} 
\noindent 2) Symmetric bidirectional DNA synthesis initiated (nucleated) at the origin corresponds to solid-domain growth.  \vspace*{1em} 
\noindent 3) Coalescence in crystal growth is analogous to multiple dispersed 
sites of replicating DNA (replication fork) that advance from opposite
directions until they merge.  \vspace*{1em}
\subsection*{Simple version of the KJMA model for DNA replication}
	
In the simplest form of the KJMA model, solids nucleate anywhere 
in the liquid, with equal probability for all spatial locations 
(``homogeneous nucleation''), although it is straightforward to
describe nucleation at pre-specified sites (``heterogeneous 
nucleation''), which would correspond to a case where replication 
origins are specified by fixed genetic sites along the genome.  
Once a solid domain has been nucleated, 
it grows out as a sphere at constant velocity $v$.  
When two solid domains impinge, growth ceases at the point of contact,
while continuing elsewhere.  KJMA used elementary methods to calculate 
quantities such as $f(\tau)$, the fraction of the volume that has 
crystallized by time $(\tau)$.  Much later, more sophisticated 
methods were developed to describe the detailed statistics of 
domain sizes and spacings.\cite{Sekimoto, Ben-Naim}
	
DNA replication, of course, corresponds to one-dimensional crystal 
growth; the shape in three dimensions of the 
one-dimensional DNA strand does not directly affect the kinetics 
modeling. (In the model, replication is one dimensional along the 
DNA.  The configuration of DNA in three dimensions is not directly 
relevant to the model but can enter
indirectly via the nucleation function $I(x, \tau)$.  For example, if, 
for steric reasons, certain regions of the DNA are inaccessible to 
replication factories, those regions would have a lower (or even zero) 
value of $I$.)  The one-dimensional version of the KJMA model 
assumes that domains 
grow out at velocity $v$, assumed to remain constant.  The nucleation 
rate $I(x,\tau) = I_0$ is defined to be the probability of domain formation 
per unit length of unreplicated DNA per unit time, 
at the position $x$ and time $\tau$. Following the analogy to the 
one-dimensional KJMA model, we can calculate 
the kinetics of DNA replication during $S$-phase. This requires determining 
the fraction of the genome $f(\tau)$ that has already been 
replicated at any given moment during $S$-phase. 
One finds
\begin{equation}
     f(\tau) =  1 - e^{-I_{0}v\tau^{2}} ,
\label{eq:ft}
\end{equation}
which defines a sigmoidal curve. (Eq. \ref{eq:ft} 
assumes an infinite genome length.  The relative importance of 
the finite size of chromosomes is set by the ratio 
(fork velocity * duration of $S$-phase) / chromosome 
length (Cahn, 1996).  In the case of the experiment analyzed in 
this paper, this ratio is $\approx$ 10 bases/sec * 1000 sec 
/ $10^7$ bases/chromosome $\approx 10^{-3}$, which we neglect.)
	
A more complete description of replication kinetics requires detailed
analysis of different statistical quantities, including measurements 
made on replicated regions (eyes), unreplicated regions (holes), and
eye-to-eye sizes (the eye-to-eye size is defined as the length between
the center of one eye and the center of a neighboring eye.)
The probability distributions may be expressed 
as functions either of time $\tau$ or replicated fraction $f$.
For example, the distribution of holes of size $\ell$ at time $\tau$, 
$\rho_h (\ell,\tau)$ can be derived by a simple extension of the 
argument leading to Eq. \ref{eq:ft}:
\begin{equation}
     \rho_h (\ell, \tau) = I_0\tau \cdot e^{-I_0\tau\ell} .
\label{meanhole}
\end{equation}
From Eq.~\ref{meanhole}, the mean size of holes at time 
$\tau$ is 
\begin{equation}
     \ell_h (\tau) = {1 \over I_0\tau} .
\label{eq:i-tau}
\end{equation}
	
Determining the probability distributions of replicated lengths (eye sizes)
is complicated because a given replicated 
length may come from  a single origin or it may result from the 
merger of two or more replicated regions.  Thus, one must calculate in 
effect an infinite number of probabilities; by contrast, holes of a given 
length arise in only one way.\cite{Ben-Naim}  
One can nonetheless derive a simple expression for $\ell_i (\tau)$, 
the mean replicated length at time $\tau$, from a 
{\it ``mean-field''} hypothesis\cite{Plischke_Bergersen}: 
the probability distribution of a given replicated length is 
assumed to be independent of the actual size of its neighbor.  
One can show that this mean-field hypothesis must always be true in 
one-dimensional growth problems, but not necessarily in the ordinary 
three-dimensional setting of crystal growth.  In particular, 
if $I(\tau)$ depends on space, one expects correlations to be important.  
Using the mean-field hypothesis, we find
\begin{equation}
     \ell_i(\tau) = \ell_h(\tau) {f \over 1 - f} = {e^{I_0 v\tau^2} - 1 
                     \over I_0 \tau}
\label{eq:ell-i-tau}
\end{equation}
and
\begin{equation}
     \ell_{i2i}(\tau) = \ell_i(\tau) + \ell_h(\tau) = {\ell_h(\tau) \over 1-f} 
                      = {e^{I_0 v\tau^2} \over I_0 \tau} .
\label{eq:ell-i2i-tau}
\end{equation}
The minimum average eye-to-eye size, obtained by differentiating Eq. \ref{eq:ell-i2i-tau}, is $\ell_{i2i}^{*} = \sqrt{2e} \cdot \sqrt{v/I_0}$.  These expressions for $\ell_i(\tau)$ and $\ell_{i2i}(\tau)$ 
allow one to collapse the experimental observations of 
$\ell_h$, $\ell_i$, and $\ell_{i2i}$ (the 
mean eye-to-eye separation) onto a single curve.  
(See Fig. \ref{fig:mean-curves}D, below.)
	
Finally, we can calculate the average distance between origins of 
replication that were activated at different times during the 
replication process, which is just the inverse of $I_{tot}$, the 
time-integrated nucleation probability per unit length:
\begin{equation}
     \ell_0 \equiv I_{tot}^{-1} 
     = {2\over\sqrt{\pi}} \cdot \sqrt{v\over I_0}
\end{equation}
The last expression shows that, as might have been guessed by 
dimensional analysis of the model parameters ($I_0$ and $v$), 
the basic length scale in the model is set by 
$\ell^* \equiv \sqrt{v/I_0}$.  Note that because initiation in the model is occurring throughout $S$-phase, the minimum eye-to-eye distance $\ell_{i2i\_min}$ is not the same as the average separation between origins, $\ell_{0}$. For this simple case, $\ell_{i2i\_min}/ \ell_{0} = \sqrt{e \pi / 2} \approx 2.1$.
\subsection*{Generalizations of the KJMA model}
	
Based on the specific results of the {\it Xenopus} experiments 
discussed above, we generalized the simple version of the KJMA model
in several ways:
	The first generalization is to allow for arbitrary $I(\tau)$.
Eq. \ref{eq:ft} then becomes
\begin{equation}
     f(\tau) = 1 - e^{-g(\tau)}~~{\rm with}~~g(\tau) = {2v \int_0^\tau I(\tau')(\tau-\tau') \, d\tau'} ,  
\label{eq:ft-general}
\end{equation}
and, similarly, Eq. \ref{eq:i-tau} becomes 
\begin{equation}
     \ell_h(\tau) = \left[{\int_0^\tau I(\tau') \, d\tau'} \right]^{-1} . 
\label{eq:ell-h-tau-general}
\end{equation}
The other mean lengths, $\ell_i(\tau)$ and $\ell_{i2i}(\tau)$, 
continue to be related to $\ell_h(\tau)$ by the general 
expressions given in Eqs. \ref{eq:ell-i-tau} and \ref{eq:ell-i2i-tau}. 
In the experiment, one measures $\ell_h$, $\ell_i$, and $\ell_{i2i}$ as 
functions of both $\tau$ and $f$.  (Because of the start-time 
ambiguity, the $f$ data are easier to interpret.)  
The goal is to invert this data to find $I(\tau)$.  
Using Eqs. \ref{eq:ft-general} and \ref{eq:ell-h-tau-general}, we find
\begin{equation}
     \tau(f) = {1\over 2v} \int_0^f \ell_{i2i}(f') \, df' 
      = {1\over 2v} \int_0^f {{\ell_h(f')} \over {1-f'}} \, df' .
\end{equation}
Because $\tau(f)$ increases monotonically, one can numerically invert
it to find $f(\tau)$.  From $f(\tau)$, one can derive all quantities 
of interest, including $I(\tau)$.
	
The starting time distribution $\phi(t)$ can be deduced looking at 
each molecular fragment, measuring its replication fraction $f$, and 
extrapolating back to a starting time using the experimentally 
determined $f(\tau)$ curve.  (Fragments that are fully replicated ($f 
= 1$ are excluded.)  The starting times are then binned to give 
$\phi(t)$ directly.

\subsection*{Monte-Carlo simulations}
	
We wrote a Monte-Carlo simulation using the programming language of Igor 
Pro (Wavemetrics) to test various experimental effects that were 
difficult to model analytically.  These 
included the effects of finite sampling of DNA fragments (on average, 190
molecules per time point), the finite optical resolution of the scanned 
images, and -- most
important -- the effect of the finite size of the combed DNA fragments. 
The size of each molecular fragment in the simulation was drawn randomly 
from an estimate of the actual size distribution of the experimental data.
This distribution was approximately log-normal, with an
average length of 102 kb. and a standard deviation of 75 kb.
	
In the simulations, we consider each DNA molecular fragment as a 
one-dimensional lattice, and each lattice site is updated with a 
timestep $\Delta t$ = 0.2 min.  An origin is initiated (lattice site 
changed from 0 to 1) with a probability determined by the initiation 
rate $I(\tau)$. Once an origin has been initiated, replication forks 
grow bidirectionally at a constant rate $v$. The
natural size of lattice then would be $v \Delta t$, which is 123 bp 
for the measured fork velocity $v = 615$ bp/min and chosen time step 
$\Delta t$.  The lattice scale is then roughly the size of origin 
recognition complex proteins.  We sample the simulation results at 
the same time points as the actual experiments ($t$ = 25, 29, 32, 35,
39, 45 min.)  Each sampled molecule is cut at random site to simulate 
the combing process.  The lattice is then ``coarse grained'' by 
averaging over approximately four pixels.  The coarse lattice length 
scale is then 0.24 $\mu$m, which roughly corresponds to that of the 
scanned optical images.  Finally, the coarse-grained fragments were 
analyzed to compile statistics concerning replicon sizes, eye-to-eye 
sizes, etc. that were directly compared to experimental data.
	
In a first version of the simulation, the lattice was directly 
simulated using a vector with one element for each lattice site.  In a 
more refined version of the simulation, we noted only the position of 
the replication forks, which greatly increased the speed of the 
simulations.
	
We also used the simulation to test a previous algorithm for 
extracting $I(f)$, the initiation rate as a function of overall 
replication fraction.  The previous algorithm\cite{Herrick_JMB, Marheineke}  looked for small replicated 
regions and extrapolated back to an assumed initiation time.  We 
tested this algorithm using our Monte-Carlo analysis and found 
significant bias in the inferred $I(f)$, while the algorithms we 
introduce here showed no such bias.

\subsection*{Parameter extraction from data}
	
We extracted data from both the real experiments and the Monte-Carlo 
simulations by a global least-squares fit that took into account 
simultaneously the different data collected (i.e., the different 
curves in Figs.~\ref{fig:rho} and \ref{fig:mean-curves}).  As discussed above, 
we fit a two-segment straight line to the $I(\tau)$ curve extracted 
directly from the data for analytic simplicity.  Assuming this form 
for $I(\tau)$, we derive explicit formulae for the curves in 
Figs.~\ref{fig:rho} and \ref{fig:mean-curves}.  
	
The finite size of the molecular fragments studied ($102 \pm 75$ kb) causes 
systematic deviation from the ``infinite-length'' formulae we can 
calculate.  Such deviations could be detected using the Monte-Carlo 
simulations by comparing the extracted values of parameters with those 
input.  The deviations show themselves mainly in two settings:
First, whenever the mean length of holes, eyes, or eye-to-eye 
distances approaches the mean segment length, the observed mean 
lengths will be systematically too small because the larger end of the 
experimental distributions is cut off by the finite fragment length.  
We dealt with this complication by restricting our fit to areas where 
the mean length being measured is less than 10\% of the mean 
fragment size.  The second complication is that the inferred fork 
velocity is systematically reduced (by about 5\% for the fragment 
size in the experiments analyzed here).  We 
measured this bias using the Monte-Carlo simulations and then corrected
the ``raw'' fork velocity that is given by our least-squares fits. 
	
One further subtle point in a global fit is the relative weighting to 
be given to the data in the $\rho(f)$ curves (Fig.~\ref{fig:rho}) 
relative to the data in the mean-value curves (Fig.~\ref{fig:mean-curves}).  We 
estimated the weights using the boot-strap method.\cite{Press}  In a similar spirit, we used repeated Monte-Carlo simulations 
to estimate statistical errors in experimentally extracted quantities.

\section*{Acknowledgments}
	
We thank M.~Wortis and B.-Y.~Ha for helpful comments and insights.  
This work was supported by grants from the Fondation de France, NSERC, and NIH.


\begin{references}

\bibitem{Hand} Hand, R. (1978). Eukaryotic DNA: organization of the 
genome for replication.  {\it Cell} {\bf 15}, 317--325.

\bibitem{Friedman} Friedman, K. L., Brewer, B.~J. \& Fangman, W.~L.
(1997).  Replication profile of {\it Saccharomyces cerevisiae} 
chromosome VI.  {\it Genes to Cells} {\bf 2}, 667--678.

\bibitem{Cairns} Cairns, J.  (1963).  The Chromosome of {\it E. coli.}  In 
{\it Cold Spring Harbor Symposia on Quantitative Biology} {\bf 28}, 43--46.

\bibitem{Shivashankar} Shivashankar, G.~V., Feingold, M., Krichevsky, O. \& Libchaber, 
A.  (1999).  RecA polymerization on
double-stranded DNA by using single-molecule manipulation:
The role of ATP hydrolysis.  {\it Proc. Natl. Acad. Sci. USA} {\bf 
96}, 7916--7921.  

\bibitem{Pierron} Pierron, G. \& Benard, M.  (1996).  DNA Replication in Physarum. 
{\it In} DNA Replication in Eukaryotic Cells. M.~DePamphilis, ed.  
Cold Spring Harbor Laboratory Press, Cold Spring Harbor.  933--946.

\bibitem{Walter_Newport} Walter, J. \& Newport, J.~W.  (1997).  Regulation of replicon size in {\it Xenopus} egg extracts. {\it Science} {\bf 275}, 993--995.

\bibitem{Hyrien_Mechali} Hyrien, O. \& Mechali, M.  (1993).  Chromosomal replication 
initiates and terminates at random sequences but at regular intervals 
in the ribosomal DNA of Xenopus early embryos. {\it EMBO J.} {\bf 
12}, 4511--4520.

\bibitem{Coverly_Laskey} Coverley, D. \& Laskey, R.~A. (1994).  Regulation of 
eukaryotic DNA replication. {\it Ann. Rev. Biochem.} {\bf 63}, 745--776.

\bibitem{Blow_Chong} Blow, J.~J.\& Chong, J.~P.  (1996).  DNA replication 
in {\it Xenopus}.  In {\it DNA Replication in Eukaryotic Cells}.
Cold Spring Harbor Laboratory Press, Cold Spring Harbor, 971--982.

\bibitem{Shinomiya} Shinomiya, T. \& Ina, S.  (1991).  Analysis of chromosomal replicons 
in early embryos of Drosophila melanogaster by two-dimensional gel
electrophoresis.  {\it Nucleic Acids Research} {\bf 19}, 3935--3941.

\bibitem{Brewer_Fangman} Brewer, B.~J.\& Fangman, W.~L.  (1993).
Initiation at Closely Spaced Replication Origins in a Yeast Chromosome. 
{\it Science} {\bf 262}, 1728--1731.

\bibitem{Gomez} Gomez, M., and F.~Antequera.  (1999).  Organization of DNA replication 
origins in the fission yeast genome.  {\it EMBO J.}  18:5683--5690.

\bibitem{Herrick_JMB} Herrick, J., Stanislawski, P., Hyrien, O. \& Bensimon, A. (2000).  
A novel mechanism regulating DNA replication in 
{\it Xenopus laevis} egg extracts. {\it J. Mol. Biol.} {\bf 300}, 1133--1142.

\bibitem{Lucas} Lucas, I., Chevrier-Miller, M., Sogo, J.~M. \& Hyrien, O.  (2000).  
Mechanisms Ensuring Rapid and Complete DNA Replication 
Despite Random Initiation in Xenopus Early Embryos. 
{\it J. Mol. Biol.} {\bf 296}, 769--786.

\bibitem{Bensimon} Bensimon A., Simon A., Chiffaudel, A., Croquette, V., Heslot, F. \& 
Bensimon, D.  (1994).  Alignment and sensitive 
detection of DNA by a moving interface. {\it Science} {\bf 265}, 2096--2098.

\bibitem{Michalet} Michalet X., Ekong, R., Fougerousse, F., Rousseaux, S., Shurra, C., 
Hornigold, N., van Slegtenhorst, M., Wolfe, J., Povey, S., Beckmann, 
J.~S.\& Bensimon, A.  (1997).  Dynamic molecular combing: 
stretching the whole human genome for high-resolution studies.  
{\it Science} {\bf 277}, 1518--1523.

\bibitem{Herrick_PNAS} Herrick, J., Michalet, X., Conti, C., Shurra, C. \& Bensimon, A.  (2000).  
Quantifying single gene copy number by measuring fluorescent probe 
lengths on combed genomic DNA. {\it Proc. Natl. Acad. Sci. USA} {\bf 
97}, 222--227.

\bibitem{Huberman} Huberman, J.~A. \& Riggs, A.~D.  (1966).  Autoradiography of chromosomal 
DNA fibers from Chinese hamster cells.
{\it Proc. Natl. Acad. Sci. USA} {\bf 55}, 599--606.

\bibitem{Jasny} Jasny, B.~R., \& Tamm, I.  (1979).  Temporal organization of replication 
in DNA fibers of mammalian cells.  {\it J. Cell Biol.}  {\bf81}, 692--697.

\bibitem{Jackson_Pombo} Jackson, D.~A. \& Pombo, A.  (1998).
Replication Clusters Are Stable Units of Chromosome Structure: 
Evidence That Nuclear Organization Contributes to the Efficient 
Activation and Propagation of S Phase in Human Cells.  
{\it J. Cell Biol.}  {\bf 140}, 1285--1295.

\bibitem{Blumenthal} Blumenthal, A.~B., Kriegstein, H.~J. \& Hogness, D.~S.  (1974).  
The units of DNA replication in {\it Drosophila 
melanogaster} chromosomes.  In {\it Cold Spring Harbor Symposia on 
Quantitative Biology} {\bf 38}, 205--223.

\bibitem{Blow_Laskey} Blow, J.~J. \& Laskey, R.~A. (1986).  Initiation of DNA 
replication in nuclei and purified DNA by a cell-free extract of {\it 
Xenopus} eggs.  {\it Cell} {\bf 47}, 577--587.

\bibitem{Blow_Watson} Blow, J.~J. \& Watson, J.~V.  (1987).  Nuclei act as 
independent and integrated units of replication in a {\it Xenopus} 
cell-free DNA replication system.  {\it Embo J.} {\bf 6}, 1997--2002.

\bibitem{Wu} Wu, J.~R., Yu, G. \& Gilbert, D.~M.  (1997). 
Origin-specific initiation of mammalian nuclear DNA replication in a 
{\it Xenopus} cell-free system. {\it Methods} {\bf 13}, 313--324.

\bibitem{Simon} Simon, I., Tenzen, T., Reubinoff, B.~E., Hillman, D., McCarrey, 
J.~R. \& Cedar, H.  (1999).  Asynchronous replication of 
imprinted genes is established in the gametes and maintained during 
development. {\it Nature} {\bf 401}, 929--932.

\bibitem{Pasero} Pasero, P. \& Schwob, E.  (2000).  Think global, act local -- 
how to regulate S phase from individual replication origins. 
{\it Current Opinion in Genetics and Development} {\bf 10}, 178--186.

\bibitem{Mills} Mills, A.~D., Blow, J.~J., White, J.~G., Amos, W.~B., Wilcock, D. \&
Laskey, R.~A.  (1989).  Replication occurs at discrete 
foci spaced throughout nuclei replicating in vitro. {\it J. Cell Sci.} 
{\bf 94}, 471--477.

\bibitem{Mahbubani} Mahbubani, H.~M., Paull, T., Elder, J.~K. \& Blow, J.~J.  (1992).  
DNA replication initiates at multiple sites on plasmid DNA in 
{\it Xenopus} egg extracts. {\it Nucl. Acids Res.} {\bf 20}, 1457--1462.

\bibitem{Lu} Lu, Z.~H., Sittman, D.~B., Romanowski, P \& Leno, G.~H.
(1998).  Histone H1 reduces the frequency of initiation in {\it Xenopus} 
egg extract by limiting the assembly of prereplication complexes on 
sperm chromatin. {\it Mol. Biol. of the Cell} {\bf 9}, 1163--1176. 

\bibitem{Buongiorno-Nardelli} Buongiorno-Nardelli, M., Michelli, G., Carri, M.~T. \& Marilley, M.  (1982).  A relationship between replicon size and 
supercoiled loop domains in the eukaryotic genome.  {\it Nature}
{\bf 298}, 100--102.  

\bibitem{Laskey} Laskey, R.~A.  (1985).  Chromosome replication in early 
development of {\it Xenopus laevis}.  
{\it J. Embryology \& Experimental Morphology} {\bf 89}, Suppl. 285--296.

\bibitem{Berezney} Berezney, R., Dubey, D.~D. \& Huberman, J.~A. (2000).  Heterogeneity
of eukaryotic replicons, replicon clusters, and replication foci.
{\it Chromosoma} {\bf 108}, 471--484.

\bibitem{Raghuraman} Raghuraman, M.~K., Winzeler, E.~A., Collingwood, D., Hunt, S., 
Wodlicka, L., Conway, A., Lockhart, D.~J., Davis, R.~W., Brewer, 
B.~J. \& Fangman, W.~L.  (2001).  Replication dynamics of the yeast 
genome.  {\it Science} {\bf 294}, 115--121.

\bibitem{Gilbert} Gilbert, D.~M. (2001).  Making sense of eukaryotic DNA replication origins.
{\it Science} {\bf 294}, 96--100.

\bibitem{Rowles} Rowles, A., Chong, J.~P., Brown, L., Howell, M., Evan, G.~I. \& Blow, J.~J. (1996). Interaction between the origin recognition complex and the replication licensing system in {\it Xenopus}. {\it Cell} {\bf 87}, 287--296.

\bibitem{Blow_etal} Blow, J. J., Gillespie, P. J., Francis, D. \& Jackson, D. A. (2001).
Replication origins in {\it Xenopus} egg extract are 5Ð15 kilobases apart
and are activated in clusters that fire at different times.  {\it J. 
Cell Biol.} {\bf 152}, 15--25.

\bibitem{Kolmogorov} Kolmogorov, A. N.  (1937).  On the statistical 
theory of crystallization in metals.  {\it Izv. Akad. 
Nauk SSSR, Ser. Fiz.} {\bf 1}, 355--359.

\bibitem{Johnson_Mehl} Johnson,  W. A \& Mehl, P.~A.  (1939).
{\it Trans. AIMME.} {\bf 135}, 416--442.  Discussion 442--458.

\bibitem{Avrami} Avrami, M.  (1939).  Kinetics of Phase Change. I.  
General theory. {\it J. Chem. Phys.}  {\bf 7}, 1103--1112. (1940).  Kinetics of 
Phase Change. II.  Transformation-time relations for random distribution 
of nuclei. {\it Ibid.}  {\bf 8}, 212--224.  (1941).  Kinetics of phase 
change III.  Granulation, phase change, and microstructure.   {\it 
Ibid.}  {\bf 9}, 177--184.

\bibitem{Christian} Christian, J. W.  (1981).  {\it The Theory of Phase 
Transformations in Metals and Alloys, Part I:  Equilibrium and 
General Kinetic Theory}.  Pergamon Press, New York.

\bibitem{Sekimoto} Sekimoto, K. (1991).  Evolution of the domain 
structure during the nucleation-and-growth process with non-conserved 
order parameter.  {\it Int. J. Mod. Phys. B} {\bf 5}, 1843--1869.

\bibitem{Ben-Naim} Ben-Naim, E. \& Krapivsky, P.~L.  (1996).  
Nucleation and growth in one dimension.  {\it Phys. Rev. E}
{\bf 54}, 3562--3568.

\bibitem{Plischke_Bergersen} Plischke, M. \& Bergersen, B.  (1994). 
{\it Equilibrium Statistical Physics}, 2nd ed., Ch. 3.  
World Scientific, Singapore.

\bibitem{Cahn} Cahn, J.~W.  (1996).  Johnson-Mehl-Avrami Kinetics 
on a Finite Growing Domain with Time and Position Dependent Nucleation 
and Growth Rates.  {\it Mat. Res. Soc. Symp. Proc.}  {\bf 398}, 425--438. 

\bibitem{Marheineke} Marheineke, K. \& Hyrien O.  (2001).  Aphidicolin triggers a block 
to replication origin firing in {\it Xenopus} egg extracts.
{\it J. Biol. Chem.} {\bf 276}, 17092--17100. 

\bibitem{Press} Press, W.~H., Teukolsky, S. A., Vetterling, W.~T. \& Flannery, B.~P. 
(1992).  {\it Numerical Recipes in C:  The Art of Scientific 
Computing}, 2nd ed., Ch. 15.  Cambridge University Press, Cambridge.

\end{references}
\end{document}